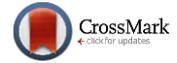

# Supporting software documentation with source code summarization

Ra'Fat Al-Msie'deen *, Anas H. Blasi

*Department of Computer Information Systems, Faculty of IT, Mutah University, P.O. Box 7, Mutah 61710, Karak, Jordan*

A B S T R A C T

Source code summarization is a process of generating summaries that describe software code, the majority of source code summarization usually generated manually, where the summaries are written by software developers. Recently, new automated approaches are becoming more useful. These approaches have been found to be effective in some cases. The main weaknesses of these approaches are that they never exploit code dependencies and summarize either the software classes or methods but not both. This paper proposes a source code summarization approach (Suncode) that produces a short description for each class and method in the software system. To validate the approach, it has been applied to several case studies. Moreover, the generated summaries are compared to summaries that written by human experts and to summaries that written by a state-of-the-art solution. Results of this paper found that Suncode summaries provide better information about code dependencies comparing with other studies. In addition, Suncode summaries can improve and support the current software documentation. The results found that manually written summaries were more precise and short as well.

## 1. Introduction

Software developers are concerning about good software documentation that involves summaries of its code to efficiently comprehend software code (Forward and Lethbridge, 2002). In fact, Code summary is a short description about a particular identifier of software (e.g., class and method). For example, the class summaries in the JavaDocs which written by human experts. The manually written summaries such as JavaDocs help software developers to comprehend small sections of code (e.g., method) with no need to comprehend the whole software code (Roehm et al., 2012). The process of writing the source code summaries is expensive and may be incomplete. In addition, code summaries may become out dated as the code evolves (McBurney and McMillan, 2016a), and may be inaccurate as they were written by human experts, who may be influenced by stress and other factors (McBurney and McMillan, 2016b). The manually written summaries did not contain any contextual information, which shows how the software identifier interacts with other ones. Software summaries reflect the features (Al-Msie'deen et al., 2014b) or functionalities it offers to the user.

Program maintenance is the most expensive and time-consuming phase in the program life cycle (Yau and Collofello, 1980). If there is no previous information about software code, programmers can have problem understanding which section of code need change (Dave et al., 2014). Software developers are making a lot of efforts to comprehend software source code. In this case, the code summaries are very important for software developers in order to document legacy software and comprehend its code. Software developers face several problems with legacy software systems due to absence of software documentation.

Source code summarization aims to create documentation for legacy software by analyzing its code. Legacy programs are usually documented based on its available artifacts such as: source code and design documents like use-case diagram (Al-Msie'deen et al., 2014c). Several studies have found that code summaries help software developers comprehend code better (McBurney and McMillan, 2014). The difficulties in writing summaries of program code have encouraged researchers to design numerous approaches as alternatives to writing code descriptions manually (cf. Section 2). This paper proposes an automatic approach

---

* Corresponding Author.
Email Address: rafatalmsiedeen@mutah.edu.jo (R. Al-Msie'deen)
https://doi.org/10.21833/ijaas.2019.01.008
Corresponding author's ORCID profile:
https://orcid.org/0000-0002-9559-2293

(Suncode†) to produce summaries of software classes and methods.

The source code summarization process is a critical issue in the software engineering domain, so numerous methods have been suggested to summarize the software code (cf. Section 2). The source code summarization process is a tedious job and costly procedure but the majority of source code summarization generators are still manual (McBurney and McMillan, 2014) which produced by human experts. However, these summaries often incomplete (Moreno et al., 2013a) and must be updated constantly (Shi et al., 2011).

Human experts spend a lot of time reading and browsing the software source code (Ko et al., 2006; LaToza et al., 2006). Thus, there is a need to propose Suncode approach to summarize the software code and overcome the limitations of existing approaches. In the context of this paper, summary is a text generated from software source code contains important information in the original text (i.e., source code), moreover, the summary text does not exceed half of the original text (Radev et al., 2002).

Suncode is a unique approach, where it summarizes the software classes and methods. Also, it focuses on the context of the identifier. In this paper, context means how the software identifier (e.g., class, attribute and method) interacts with other ones via code dependencies such as: inheritance, method invocation and attribute access. Suncode produces a readable English summary of the context for each class and method in the software system.

Several approaches have been proposed to summarize the software source code. One approach presented by Haiduc et al. (2010b) returns a list of keywords from the method. A different method presented by Sridhara et al. (2010) produces descriptive summary comments for software methods. Moreno et al. (2013b) presented an approach to summarize the software classes. McBurney and McMillan (McBurney and McMillan, 2014) suggested an approach to generate summaries of the context of Java methods.

This paper proposes an approach to summarize the software classes and methods based on their textual and structural information (Suncode accepts as input software code and produces as outputs a set of summaries). Based on the static code analysis, Suncode extracts the software source code. Then, Suncode generates rapid summary messages (McBurney and McMillan, 2014) for each class and method and, at last, Suncode aggregates all rapid summary messages into a single document.

Suncode approach is detailed in this paper as follows: Section 2 discusses the related work relevant to Suncode contributions. Section 3 shows an overview of Suncode. Section 4 presents the source code summarization process step-by-step. Section 5 describes the experiments that were conducted to validate Suncode proposal, while section 6 concludes and provides perspectives for this work.

## 2. Related works

This section presents the related work closest to Suncode approach. McBurney and McMillan (2016b, 2014) presented a source code summarization method that generates descriptions of the context of Java methods – that is, where the method is called. They use static code analysis. In their work, the context of the method is limited to method invocation. Suncode presents an automatic approach to summarize the software classes and methods. The context of the class is how the class interacts with other ones. While the context of method is how the method interacts with other methods and attributes. Suncode helps software developers understand the role the class (resp. method) plays in the software. Suncode uses static code analysis.

A different work was developed by Sridhara et al. (2010). Sridhara et al. (2010) presented a new method to automatically produce descriptive summary comments for software methods. Based on the method's signature and body, their comment generator found the content of the summary and produced the natural language text that summarizes the method.

Another method offered by Haiduc et al. (2010b) produced a list of keywords from the method code using the vector space model. Haiduc et al. (2010b) proposed using a vector space model method to mine important keywords from method and presented those keywords to software developers. In fact, their approach helps software developers to get some contextual information through listing the main keywords from method code.

Moreno et al. (2013a) suggested a new method to automatically produce natural language summaries for software classes by using stereotype of class. In another paper, Moreno et al. (2013b) presented an Eclipse plugin named jSummarizer‡ to produce the natural language summaries for software classes based on a stereotype of class. In their work, the produced summaries used to re-document the class code. Moreover, the extracted summaries used to help programmers to easier comprehend complex software classes.

Suncode developed to document legacy software systems. For each class and method in the software there is a summary. The produced summary for the class or method contains textual and structural information based on its code. Suncode used static code analysis to parse the software code. Suncode accepts as inputs the software code and produces a summary for each class and method in the software. Al-Msie'deen (2014) approach relied on generating rapid summary messages for each class and method in the software. The rapid summary message is a predefined template. This template is a natural language sentence describing a particular kind of rapid summary message. Each template is filled in with keywords selected from the extracted code. Suncode approach constructs summaries by combining the rapid summary messages.

---

† Suncode stands for Supporting Software Documentation with Source Code Summarization.

‡ JSummarizer: http://www.cs.wayne.edu/~severe/jsummarizer/

## 3. Approach overview

This section presents the main ideas used in Suncode. It also provides an overview of the source code summarization process. Then, it introduces the drawing shapes software that illustrates the remaining of the paper.

Various papers in the domain of software understanding show that software developers rely on good software documentation. In general, manually-written documentation is incomplete, time-consuming and it must be updated periodically (McBurney and McMillan, 2014). Thus, Suncode comes to overcome these problems regarding manually-written documentation. Suncode automatically document software code by producing a useful summary.

Suncode aims to provide software developers with meaningful summary for software classes and methods. It uses static code analysis to parse software code. The source code summarization process takes the software's code as its inputs and generates a set of summaries as its outputs. Furthermore, the code summaries are important to document and understand legacy software systems.

Suncode exploits software identifiers and code dependencies to produce a useful summary. Names of software identifiers consider as textual information. On the other hand, the dependencies between code elements consider as structural information. Software functionality is presented by its classes and methods. Thus, Suncode produces two types of summaries one for class and another for method.

The basic point that distinguishes Suncode's approach from other approaches is that it deals with the context of the class and method, which contains textual and structural information of the class and method. The context of the class being summarized includes the class signature and its body. On the other hand, the context of the method being summarized contains the method signature and its body. The context of class or method is important because it helps software developers to know the behavior and the role of the class or method in the software system.

Fig. 1 shows the source code summarization process. Suncode approach creates a summary for given software in three steps: 1) Extract the software source code using static code parser. Then, 2) Identify rapid summary messages for software classes and methods. Finally, 3) Generate English readable sentences to summarize each class and method in the software system. The summarization process is automatic; as long as the software contains classes and methods, the summaries will describe the main information of a given software class or method.

As an illustrative example, this paper considers the drawing shapes software[§]. However, this software system uses to draw several types of shapes (Al-Msie'deen, 2018; Al-Msie'deen and Blasi, 2018). This toy software used for better explanation for the remaining of this paper. Suncode only uses the software code as input for the software code summarization process.

## 4. Source code summarization process

This section presents the source code summarization process step-by-step. Suncode identifies code summaries in three steps as detailed below.

### 4.1. Extracting the software source code

Suncode uses static code analysis to parse software source code. Suncode parser[**] produces an XML file for each software system. The XML file contains main code elements such as: package, class, attribute and method. Also, it contains the main dependencies between code elements such as: inheritance, method invocation and attribute access. The extracted XML file contains textual and structural information about software code. XML format is readable by the human experts and its representation is independent of any programming language (Kanellopoulos et al., 2006).

To document the legacy software system by summarizing its code there are needs to extract main code elements and main dependencies between those elements. Fig. 2 shows the format of the XML file which uses to express the object-oriented source code.

### 4.2. Identifying rapid summary messages for each class and method

Based on the extracted XML file from the previous step, Suncode produces different types of messages that represent information about a class's/method's context. These messages are called rapid summary messages. Suncode creates six different types of rapid summary messages that represent information about a class's context. These rapid summary messages are briefly described in Table 1.

The goal of these rapid summary messages is to determine the content of each class in a software system. First, a name message gives the name of the class. The access level message is a message created to reflect the access level of the class (e.g., public). Another type of message is the package message. The idea behind a package message is to give software developers the package to which the class belongs. A fourth message type is the inheritance message. This message presents information about the class that the class inherits from. Another type of message is the attribute message. This message mentions all attributes that belong to the class. The last message type is the method message. This message serves to mention all methods that belong to the class. These messages are briefly shown in Fig. 3.

---

[§] Drawing shapes: https://sites.google.com/site/ralmsideen/tools
[**] Suncode parser https://sites.google.com/site/ralmsideen/tools

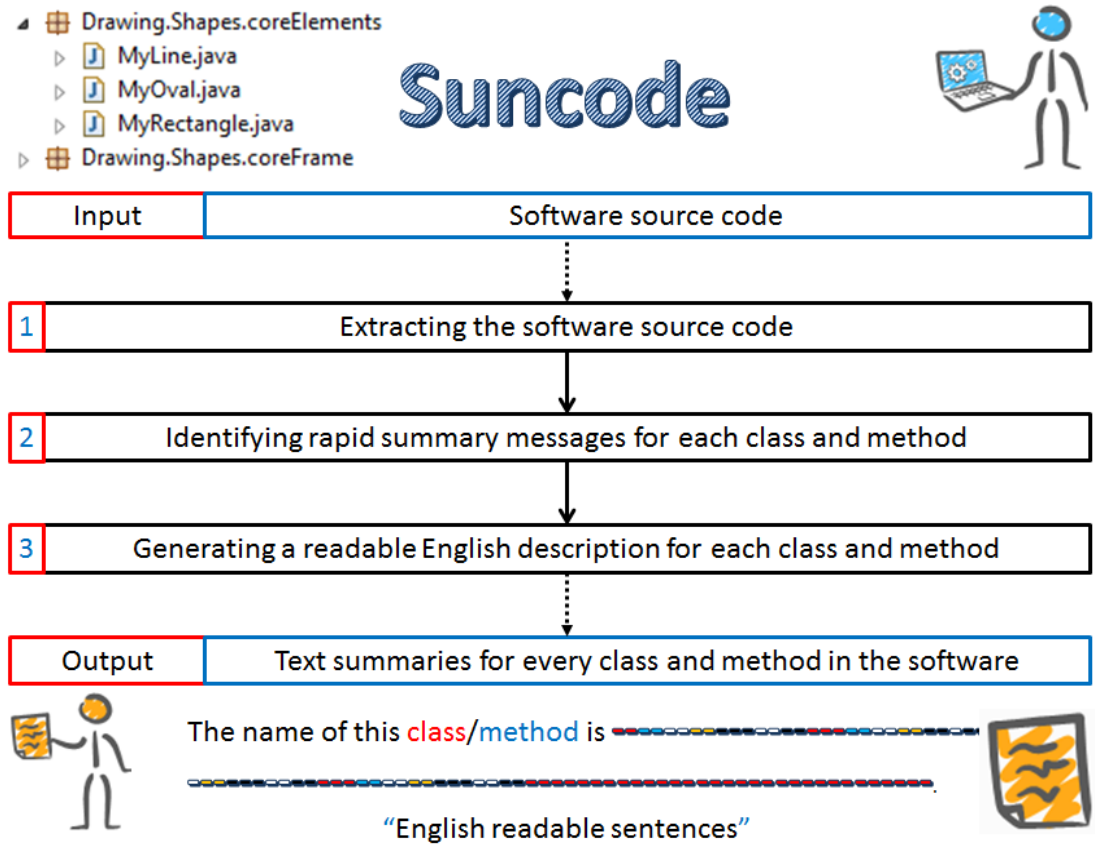

**Fig. 1:** Source code summarization process

**Fig. 2:** XML as a format of expression of object-oriented source code

Moreover, Suncode generates eight different kinds of rapid summary messages that give information about a method's context. These rapid summary messages are briefly described in Table 2.

Table 1: Rapid summary messages that Suncode creates for class's context

| # | Message type | Description |
|---|---|---|
| 1 | Name message | Gives the name of the class |
| 2 | Access level message | Gives the access level of the class |
| 3 | Package message | Gives the name of the package to which the class belongs |
| 4 | Inheritance message | Gives the name of the class that the class inherits from |
| 5 | Attribute message | Gives the names of attributes that belong to the class |
| 6 | Method message | Gives the names of methods that belong to the class |

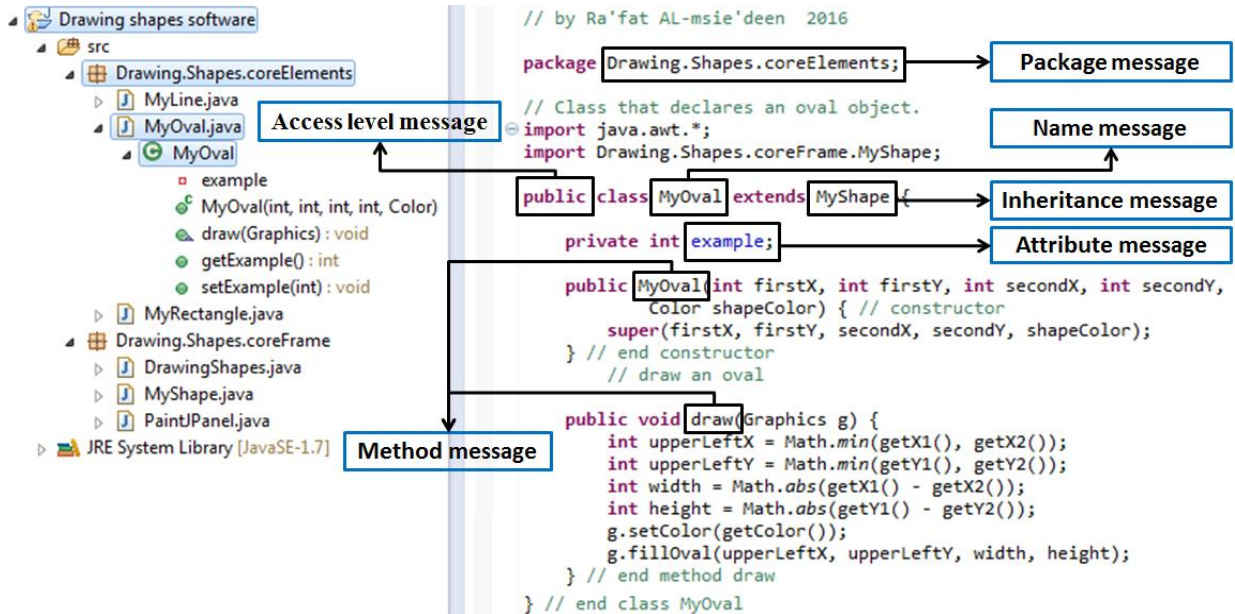

**Fig. 3:** Rapid summary messages that Suncode creates for class's context

Table 2: Rapid summary messages that Suncode creates for method's context

| # | Message type | Description |
|---|---|---|
| 1 | Name message | Gives the name of the method |
| 2 | Access level message | Gives the access level of the method |
| 3 | Return type message | Gives the return data type of the method |
| 4 | Class message | Gives the name of the class to which the method belongs |
| 5 | Parameter message | Gives the names, data types and order of the parameters |
| 6 | Variable message | Gives the names and data types of local variables |
| 7 | Access message | Gives the names of the attributes that the method accesses |
| 8 | Invocation message | Gives the names of the methods that the method calls |

The main objective of these rapid summary messages is to determine the content of each method in the software product. The first message is the name message. This message gives the name of the method. Then, the access level message aims to return the access level of method (e.g., public, private or protected). A third message type is the return type message. The return type message is a message created to give the return data type of the method. Another type of message is the class message. The idea behind a class message is to give programmers the class to which the method belongs. The parameter message is a message created to return the number of parameters, names, data types and order of the parameters in the method signature. While, the variable message is a message conveys information about the names and data types of local variables inside the method body. The access message returns the names of the attributes that the method accesses and, at last, the final message type is the invocation message. This message is used to say what methods a given method calls or invokes. The method's rapid summary messages are shown in Fig. 4.

Suncode accepts the software source code. Then, extracts the software code based on the static code analysis (Al-Msie'deen, 2015). After that, the approach identifies the rapid summary messages that summarize the software classes and methods. For example, in the drawing shapes software, the rapid summary messages for myoval class are shown in Table 3.

Table 3: An example of rapid summary messages produced by Suncode for class's context (e.g., myoval class)

| # | Message kind | Message text |
|---|---|---|
| 1 | Name message | The name of this class is MyOval. |
| 2 | Access level message | The access level for this class is public. |
| 3 | Package message | The package to which this class belongs is coreElements. |
| 4 | Inheritance message | This class inherits from the MyShape class. |
| 5 | Attribute message | This class contains the following attribute: example. |
| 6 | Method message | This class contains the following methods: MyOval and draw. |

For method's context, Suncode generates different message types to give a summary for each method in the software system. For instance, in the drawing shapes software, the rapid summary messages for main method are shown in Table 4.

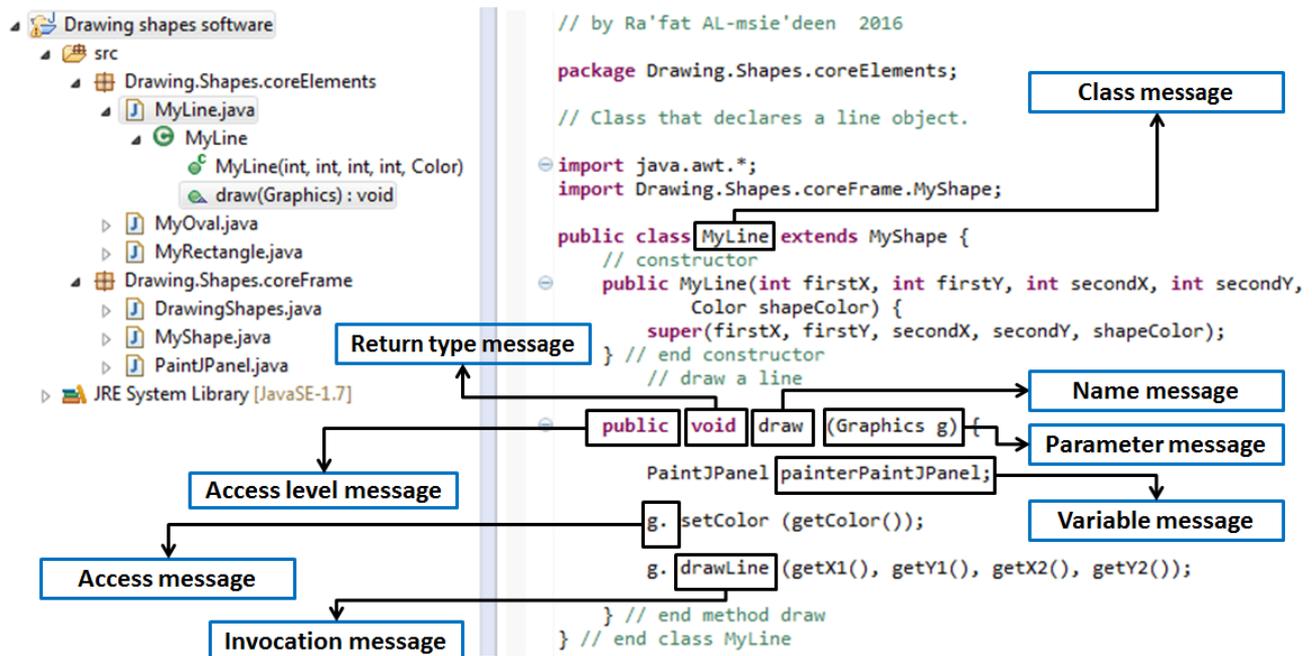

**Fig. 4:** Rapid summary messages that Suncode creates for method's context

**Table 4:** An example of rapid summary messages generated by Suncode for method's context (e.g., main method)

| # | Message kind | Message text |
|---|---|---|
| 1 | Name message | The name of this method is main. |
| 2 | Access level message | The access level for this method is public. |
| 3 | Return type message | The return data type for this method is void. |
| 4 | Class message | The class to which this method belongs is drawingShapes. |
| 5 | Parameter message | This method contains 1 parameter. This method consists of the following parameter: args and its data type is string. |
| 6 | Variable message | This method contains the following local variable: application and its data type is drawingShapes. |
| 7 | Access message | This method accesses the following attributes: application and exit_on_close. |
| 8 | Invocation message | This method invokes the following method: setDefaultCloseOperation. |

The main goal of the code summarization is to describe the source code of any software system. The functionalities of any software implemented through its classes and methods. Thus, Suncode focuses on the class and method summaries.

### 4.3. Generating a readable English description for each class and method

Suncode creates a summary of a given identifier (class or method) in three steps: a) extracting software code, then, b) generating rapid summary messages and, at last, c) aggregating all rapid summary messages into a single document. Suncode summarizes software classes and methods as readable English text and generates a document for each class and method, where messages occur in a predefined order as in Tables 1 and 2. This ordering was decided based on the importance of the messages from the author's point of view. The code summary contains textual and structural information (Al-Msie'deen et al., 2013). The main dependencies between source code elements (i.e., inheritance, method invocation and attribute access) represent structural code information, while the software identifiers and their properties (e.g., name and access level) consider as textual code information.

In the last step of the code summarization process, Suncode aggregates the rapid summary messages into a single document. For example, the summary of myoval class (Table 3) is "The name of this class is MyOval. The access level for this class is public. The package to which this class belongs is coreElements. This class inherits from the MyShape class. This class contains the following attribute: example. This class contains the following methods: MyOval and draw".

As another example, consider the main method from the drawing shapes software. The Suncode summary is "The name of this method is main. The access level for this method is public. The return data type for this method is void. The class to which this method belongs is drawingShapes. This method contains 1 parameter. This method consists of the following parameter: args and its data type is string. This method contains the following local variable: application and its data type is drawingShapes. This method accesses the following attributes: application and exit_on_close. This method invokes the following method: setDefaultCloseOperation". Moreover, the fully combined summary of draw method is shown in Fig. 5 (McBurney and McMillan, 2016a).

Software developers need an efficient method to help them to understand the legacy software code. Reading the complete code of software takes too long time. In addition, the reading of method (or class) signature does not tell us enough information about the purpose of the code. Thus, Suncode aims to create a brief description for software code by using textual and structural information in software classes and methods.

**Fig. 5:** An example of a summary generated by Suncode approach (e.g., draw method)

### 5. Experimentation

To validate Suncode, the experiments were conducted on two Java open-source software systems: i.e., NanoXML†† and ArgoUML‡‡. NanoXML software is a Java program for parsing XML file. ArgoUML is an open source tool for UML diagrams. The two software systems show different sizes: NanoXML (medium system) and ArgoUML (large system).

The different complexity levels display the scalability of Suncode to dealing with such case studies. NanoXML and ArgoUML software is well documented and their code summaries are available for comparison to Suncode summaries and validation of its proposal.

Suncode summaries are compared to summaries written by human experts such as Javadocs. Javadoc is a type of software documentation which is a summary written by the software developers. Table 5 shows the obtained summary for getResult method through Suncode approach and Javadoc.

Moreover, Suncode summaries are compared to summaries written by a state-of-the-art solution (McBurney and McMillan, 2014; 2016a). Table 6 shows the obtained summary for read method through Suncode approach and McBurney tool (McBurney and McMillan, 2016a).

Table 7 presents the extracted summary of the ArgoStatusEvent class from ArgoUML software using Suncode prototype§§. In addition to the summary from JavaDocs.

Results note that there is a difference between the summaries written by software developers, and summaries generated by Suncode approach. Comparing to manually written summaries and the state-of-the-art summaries, results found that Suncode summaries provide better contextual information. Suncode summaries can improve the current software documentation by combining Suncode summaries with other summaries. In contrast, manually written summaries were more precise and short.

Based on the Suncode results, it appears the summary is generated based on the class (resp. method) signature and its body. The extracted summary is clear, precise and brief. The extracted summary tells programmers, where the method is called. The produced summary tells software developers where the class is inherited. The summary contains too many details and helps programmers to understand what the class and method do. In addition, summaries provide helpful contextual information about the software classes and methods. The generated summaries can be used to improve existing documentation.

One threat to the validity of Suncode approach is that it considers only the Java software systems. This represents a threat to prototype validity (Al-Msie'deen, 2014) that limits Suncode implementation ability to deal only with systems that developed based on the Java language. In addition, Suncode does not include the class and method comments in the summarization process. Moreover, Suncode does not split identifier name into their constituent words where it appears as it is in the summary (should be improved by using identifier splitting algorithms).

### 6. Conclusion and future directions

This paper has presented a new approach for automatically generating summaries of software classes and methods. Suncode approach differs from other approaches in that it summarizes the software classes and methods. Moreover, Suncode exploits textual and structural information (Haiduc et al., 2010a) to summarize software classes and methods. The only input of the approach is the software code and the output is a set of English paragraphs describing software classes and methods. Suncode used rapid summary messages to identify the most important information in the class and method context. Then, Suncode aggregated messages to create an English paragraph of this context. The authors have implemented Suncode and evaluated its generated results on three case studies.

---

†† http://nanoxml.sourceforge.net/orig/index.html
‡‡ http://argouml-downloads.tigris.org/argouml-0.28.1/
§§ https://sites.google.com/site/ralmsideen/tools

**Table 5:** The extracted summary from the getResult method in NanoXML software

| | | | |
|---|---|---|---|
| Software | NanoXML | | |
| Class | StdXMLBuilder | Method | getResult ( ) |
| Method signature | public Object getResult ( ) | Method body | return this.root; |
| Javadoc summary | "This method returns the result of the building process" (NanoXML Javadocs: http://nanoxml.sourceforge.net/orig/NanoXML-2-JavaDoc/index.html) | | |
| Suncode summary | The name of this method is getResult. The access level for this method is public. The return data type for this method is object. The class to which this method belongs is StdXMLBuilder. This method accesses the following attribute: root. | | |

Results showed that all summaries were identified. The authors have compared the summaries produced from Suncode to summaries written by human specialists (e.g., Javadocs) and to summaries written by a state-of-the-art approach. The authors have found that Suncode provided better contextual information than manually written and the state-of-the-art summaries.

**Table 6:** The mined summary from the read method in NanoXML software

| | | | |
|---|---|---|---|
| Software | NanoXML | | |
| Class | StdXMLReader | Method | read ( ) |
| McBurney tool summary | "This method reads a character and returns the character. That character is used in methods that add child XML elements and attributes of XML elements. This method calls a method that skips the whitespace. This method can be used in an assignment statement; for example: char ch = reader.read();" (McBurney and McMillan, 2016a). | | |
| Suncode summary | The name of this method is read. The access level for this method is public. The return data type for this method is character. The class to which this method belongs is StdXMLReader. This method contains the following local variable: ch and its data type is int. This method accesses the following attributes: currentReader, pbReader and readers. This method invokes the following methods: read, empty, close, pop and read. | | |

Furthermore, authors can improve the current software documentation by combining Suncode summaries with the manually written summaries or the state-of-the-art summaries. In contrast, manually written summaries were more precise and short. For future work, Suncode plans to split the names of software identifiers (e.g., package, class, attribute and method) into words by using the camel-case splitting algorithm (Al-Msie'deen et al., 2014a). It also plans to use the class and method comments in the summarization process.

**Table 7:** The extracted summary from the ArgoStatusEvent class of ArgoUML software

| | |
|---|---|
| Software | ArgoUML |
| Class | ArgoStatusEvent |
| Javadoc summary | "The status event is used to notify interested parties of a status change" (ArgoUML Javadocs: http://argouml-stats.tigris.org/nonav/javadocs/javadocs-0.28/) |
| Suncode summary | The name of this class is ArgoStatusEvent. The access level for this class is public. The package to which this class belongs is org.argouml.application.events. This class inherits from the ArgoEvent class. This class contains the following attribute: text. This class contains the following methods: ArgoStatusEvent, getEventStartRange and getText. |

## Compliance with ethical standards

## Conflict of interest

The authors declare that they have no conflict of interest.